\documentclass[prb,twocolumn,amscd,amsmath,amssymb,verbatim]{revtex4}

\usepackage{graphicx}
\usepackage[dvips]{hyperref}
\usepackage[TS1,OT1,T1]{fontenc}

\begin{document}
\title{Neutron scattering and scaling behavior in URu$_2$Zn$_{20}$ and YbFe$_2$Zn$_{20}$}

\author{C. H. Wang$^{1,2}$, A. D. Christianson$^3$, J. M. Lawrence$^1$, E. D. Bauer$^2$, E. A.
Goremychkin$^{4}$, A. I. Kolesnikov$^{3}$, F. Trouw$^2$, F.
Ronning$^2$, J. D. Thompson$^2$, M.D. Lumsden$^3$, N. Ni$^5$, E. D.
Mun$^5$, S. Jia$^5$, P. C. Canfield$^5$, Y. Qiu$^{6,7}$ and J. R. D.
Copley$^6$ }
\affiliation{$^1$University of California, Irvine, California 92697, USA\\
  $^2$Los Alamos National Laboratory, Los Alamos, NM 87545, USA\\
  $^3$Neutron Scattering Sciences Division, Oak Ridge National Laboratory, Oak Ridge, TN, 37831, USA\\
  $^4$Argonne National Laboratory, Argonne, IL 60439, USA\\
  $^5$Ames Laboratory, Iowa State University, Ames, IA, 50011\\
  $^6$National Institute of Standards and Technology, Gaithersburg, Maryland 20899-6102, USA\\
  $^7$Department of Materials Science and Engineering, University of Maryland, College Park, Maryland 20742, USA}

\date{\today}

\begin{abstract}

The dynamic susceptibility $\chi^{\prime\prime}(\Delta E)$, measured
by inelastic neutron scattering measurements, shows a broad peak
centered at $E_{max}$ = 16.5 meV for the cubic actinide compound
URu$_2$Zn$_{20}$ and 7 meV  at the (1/2, 1/2, 1/2) zone boundary for
the rare earth counterpart compound YbFe$_2$Zn$_{20}$. For
URu$_2$Zn$_{20}$, the low temperature susceptibility and magnetic
specific heat coefficient $\gamma = C_{mag}/T$ take the values $\chi
=$ 0.011 emu/mole and  $\gamma =$  190 mJ/mole-K$^2$ at $T$ = 2 K.
These values are roughly three times smaller, and $E_{max}$ is three
times larger, than  recently reported for the related compound
UCo$_2$Zn$_{20}$, so that $\chi$ and $\gamma$ scale inversely with
the characteristic energy for spin fluctuations, $T_{sf} =
E_{max}/k_{B}$. While $\chi(T)$, $C_{mag}(T)$, and $E_{max}$ of the
4$f$ compound YbFe$_2$Zn$_{20}$ are very well described by the Kondo
impurity model, we show that the model works poorly for
URu$_2$Zn$_{20}$ and UCo$_2$Zn$_{20}$, suggesting that the scaling
behavior of the actinide compounds arises from spin fluctuations of
$\emph{itinerant}$ 5$f$ electrons.

\end{abstract}

\vskip 15 pt

\pacs{71.27.+a, 72.15.Qm, 75.20.Hr}

\maketitle

\section{introduction}

An important property of heavy fermion (HF) materials is a scaling
law whereby the low temperature magnetic susceptibility $\chi$ and
specific heat coefficient $\gamma = C/T$ vary as 1/$T_{sf}$. Here
$k_{B} T_{sf}$ is the spin fluctuation energy scale which can be
directly observed as the maximum $E_{max}$ in the dynamic
susceptibility $\chi^{\prime\prime}(\Delta E)$, measured through
inelastic neutron scattering. Such scaling receives theoretical
justification\cite{Hewson,Bickers,Rajan,Cox} from the Anderson
impurity model (AIM), where the spin fluctuation temperature
$T_{sf}$ is identified as the Kondo temperature $T_{K}$. This model
assumes that fluctuations in $\emph{local}$ moments dominate the low
temperature ground state properties of HF materials. For 4$f$
electron rare earth HF compounds, the AIM appears to give an
excellent description of much of the experimental behavior,
including the temperature dependence of the magnetic contribution to
the specific heat $C_{mag}$, the susceptibility $\chi$, and the 4$f$
occupation number $n_{f}$, as well as the energy dependence of the
inelastic neutron scattering (INS) spectra
$\chi^{\prime\prime}(\Delta E)$ of polycrystalline
samples\cite{slowcrossover}. The theoretical
calculations\cite{Hewson,Bickers,Rajan,Cox} show that these
properties are highly dependent on the orbital degeneracy $N_J  (=
2J+1$ for rare earths). In particular, for large degeneracy ($N_J >$
2) both the calculated $\gamma(T)$ and $\chi(T)$ exhibit maxima at a
temperature $\alpha T_{K}$ where $\alpha$ is a constant that depends
on $N_J$. This kind of behavior is observed in rare earth compounds
such as YbAgCu$_4$\cite{slowcrossover},
CeIn$_{3-x}$Sn$_x$\cite{LawrenceCeInSn}, YbCuAl\cite{Newns}, and
YbFe$_2$Zn$_{20}$\cite{CanfieldPNAS}.

It is reasonable to apply the AIM, which assumes local moments, to
rare earth compounds where the 4$f$ orbitals are highly localized
and hybridize only weakly with the conduction electrons. On the
other hand, in uranium compounds, the 5$f$ orbitals are spatially
extended and form dispersive bands through strong hybridization with
the neighboring $s$, $p$, and $d$ orbitals. Photoemission
spectroscopy in 4$f$ electron systems shows clear signals from local
moment states at energies well below the Fermi level; the weak
hybridization between the $f$ electron and the conduction electron
leads to emission near the Fermi energy $\epsilon_F$ that can be
described in the context of the Anderson impurity model as a Kondo
resonance\cite{ARPESCe115}. In 5$f$ electron systems, no local
states are seen, but rather broad 5$f$ band emission is observed
near $\epsilon_F$. The Anderson lattice model is sometimes employed
to understand the $f$-derived band in actinide
systems\cite{ARPESU122} while in some systems itinerant-electron
band models are employed\cite{ARPESU}. Hence, despite the common
occurrence of scaling, we might expect differences between the
uranium and the rare-earth based heavy fermion materials in the
details of the thermodynamics and the spin fluctuations.
Nevertheless, we have recently shown\cite{BauerCo} that the actinide
compound UCo$_2$Zn$_{20}$ exhibits a maximum in the susceptibility
and a specific heat coefficient that are strikingly similar to those
seen in the rare earth compound YbFe$_2$Zn$_{20}$. It is thus of
interest to test whether a local moment AIM/Kondo description, which
has been shown to give excellent agreement with the data for the Yb
compound (see Ref. 8 and also Fig. 3 of this paper), may also be
valid for 5$f$ HF compounds.

To accomplish this, we present herein the results of INS experiments
on polycrystalline URu$_2$Zn$_{20}$ together with results for the
magnetic susceptibility and specific heat of single crystalline
samples. We also present the INS data on single crystal
YbFe$_2$Zn$_{20}$. Both compounds belong to a new family of
intermetallic compounds RX$_2$Zn$_{20}$ (R = lanthanide, Th, U; X =
transition metal)\cite{Bauer,Thiede,Goncalves,Jia,CanfieldPNAS}
which crystallize in the cubic CeCr$_2$Al$_{20}$ type structure
($Fd\overline{3}m$ space group)\cite{Thiede,Niemann}. In this
structure, every $f$-atom is surrounded by 16 zinc atoms in a nearly
spherical array of cubic site symmetry, which leads to small crystal
field splittings. Because the R-atom content is less than 5$\%$ of
the total number of atoms, and the shortest $f/f$ spacing is $\sim$
6 {\AA}, these compounds are possible candidates for studying the
Anderson impurity model in periodic $f$ electron compounds.

\section{experiment details}

The crystals were grown in zinc flux\cite{CanfieldPNAS,BauerCo}. The
magnetic susceptibility measurements were performed in a commercial
superconducting quantum interference device (SQUID) magnetometer.
The specific heat experiments were performed in a commercial
measurement system that utilizes a relaxational (time constant)
method. For URu$_2$Zn$_{20}$, we performed inelastic neutron
scattering on a 40 gram powder sample on the low resolution medium
energy chopper spectrometer (LRMECS) at IPNS, Argonne National
Laboratory, on the High-Resolution Chopper Spectrometer (Pharos) at
the Lujan center, LANSCE, at Los Alamos National Laboratory, and on
the time-of flight Disk Chopper Spectrometer (DCS) at the NIST
Center for Neutron Research. For YbFe$_2$Zn$_{20}$ the INS spectrum
was obtained for two co-aligned crystals of total mass 8.5 grams,
using the HB-3 triple-axis spectrometer at the High Flux Isotope
Reactor (HFIR) at Oak Ridge National Laboratory (ORNL); the final
energy was fixed at $E_f$ = 14.7 meV, and the scattering plane was
$(H,H,L)$. The data have been corrected for scattering from the
empty holder but have not been normalized for absolute cross
section. For the Pharos and LRMECS measurements of URu$_2$Zn$_{20}$,
we used the non-magnetic counterpart compound ThCo$_2$Zn$_{20}$ to
determine the scaling of the nonmagnetic scattering between low $Q$
and high $Q$; for YbFe$_2$Zn$_{20}$, we measured at $Q=$
(1.5,1.5,1.5) and (4.5,4.5,4.5) and assumed that the phonon
scattering scales as $Q^2$ dependence\cite{endnote}. Assuming that
the magnetic scattering scales with the $Q$-dependence of the 4$f$
or 5$f$ form factor, we subtracted the nonmagnetic component to
obtain the magnetic scattering function $S_{mag}(\Delta
E)$\cite{slowcrossover,Murani,Eugene}.

\section{results and discussion}

The magnetic susceptibility $\chi(T)$ and the specific heat $C/T$ of
URu$_2$Zn$_{20}$ are displayed in Fig. 1 and compared to the data for UCo$_2$Zn$_{20}$. Fits
of the data to a Curie-Weiss law (Fig. 1(a)) at high temperature give the effective
moments $\mu_{eff} =$ 3.61 $\mu_B$ for URu$_2$Zn$_{20}$ and 3.44 $\mu_B$ for
UCo$_2$Zn$_{20}$. The Curie-Weiss temperatures
are $\theta$ = -145 K and -65 K for the Ru and Co cases,
respectively. For URu$_2$Zn$_{20}$, the magnetic susceptibility
$\chi(T)$ increases monotonically as the temperature decreases
to the value $\chi(2K) \simeq$ 0.0111 emu/mole. At 2 K, the susceptibility
of UCo$_2$Zn$_{20}$ is about 0.0372 emu/mole, which is 3.3 times
larger than for the Ru case. The specific heat is plotted as $C/T$ vs
$T$ in Fig. 1 (b). For URu$_2$Zn$_{20}$ $C/T$ has the magnitude
$\gamma \simeq$ 190 mJ/mole-K$^2$ at 2 K. At low temperature
$C/T$ follows the $T^2$ behavior expected for a phonon contribution, which
permits the extrapolation of the Sommerfeld coefficient to the value
$\gamma \simeq$ 188 mJ/mole-K$^2$. From the inset to Fig. 1(b), it can be
seen that for UCo$_2$Zn$_{20}$, $\gamma (2K)$ is approximately
500 mJ/mole-K$^2$, while at $T_{max}$ = 4.1 K, $\gamma$ = 558 mJ/mole-K$^2$;
these values are 2.6 and 2.9 times larger than for URu$_2$Zn$_{20}$, respectively.

\begin{figure}[t]
\centering
\includegraphics[width=0.45\textwidth]{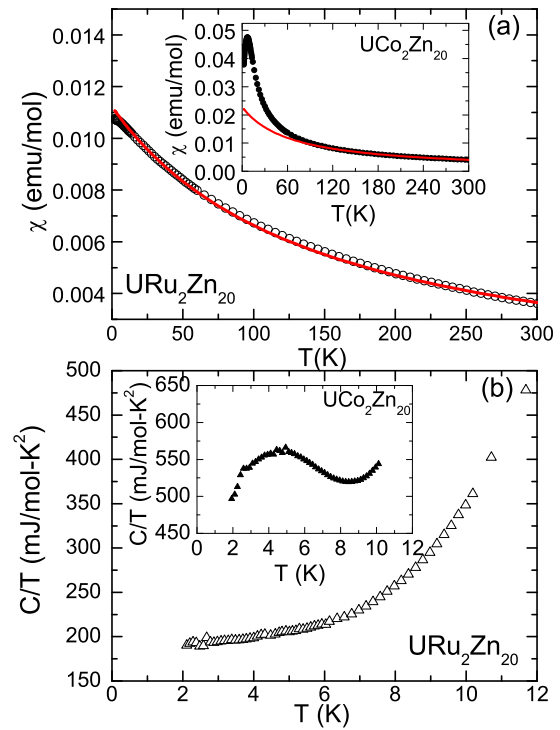}
\caption{\label{fig:1} (a) Magnetic susceptibility $\chi(T)$ for
URu$_2$Zn$_{20}$. The lines are Curie-Weiss fits. (b) Specific heat
$C/T$ vs $T$ of URu$_2$Zn$_{20}$. Insets: Susceptibility and
specific heat of UCo$_2$Zn$_{20}$; the data are from Bauer $et$
$al$\cite{BauerCo}. }\vspace*{-3.5mm}
\end{figure}

As mentioned above, the characteristic energy for spin
fluctuations can be determined from the inelastic neutron scattering experiments.
In Fig. 2 we plot the Q-averaged dynamic susceptibility
$\chi^{\prime\prime}(\Delta E)$ of URu$_2$Zn$_{20}$ as a function of energy transfer
$\Delta E$. This is determined from the scattering function through the formula
$S_{mag}=A(n(\Delta E)+1)f^2(Q)\chi^{\prime\prime} (\Delta E)$,
where $(n(\Delta E)+1)$ is the Bose factor and $f^2(Q)$ is the U 5$f$ form factor.
Both the Pharos data and the LRMECS data  for $\chi^{\prime\prime}(\Delta E)$
for URu$_2$Zn$_{20}$ exhibit broad peaks with peak position $E_{max}$ at an
energy transfer $\Delta E \simeq$ of order 16 meV. The dynamic
susceptibility $\chi^{\prime\prime}(\Delta E)$ can be fit by a Lorentzian power
function as  $\chi^{\prime\prime}(\Delta E)$= $\chi'(0) \Delta E (\Gamma /
\pi)/[(\Delta E-E_0)^2+\Gamma^2]$ with the parameters $E_0$ = 13.5 meV and
$\Gamma$= 9.5 meV, giving $E_{max}$ = 16.5 meV. As shown in the inset to
Fig. 2, for UCo$_2$Zn$_{20}$, $\chi^{\prime\prime}(\Delta E)$ shows a peak
centered near $E_{max}$ = 6 meV. Fits of this data to an
inelastic Lorentzian give  $E_0$ = 3 meV with $\Gamma$ = 5 meV, for
which $E_{max}$ = 5.8 meV. We note that these values of $E_{max}$ are nearly
equal to the values of $k_B \theta$ derived from the high temperature susceptibility;
i.e. the temperature scale for the suppression of the moment is identical to
the energy scale of the spin fluctuation.

\begin{figure}[t]
\centering
\includegraphics[width=0.45\textwidth]{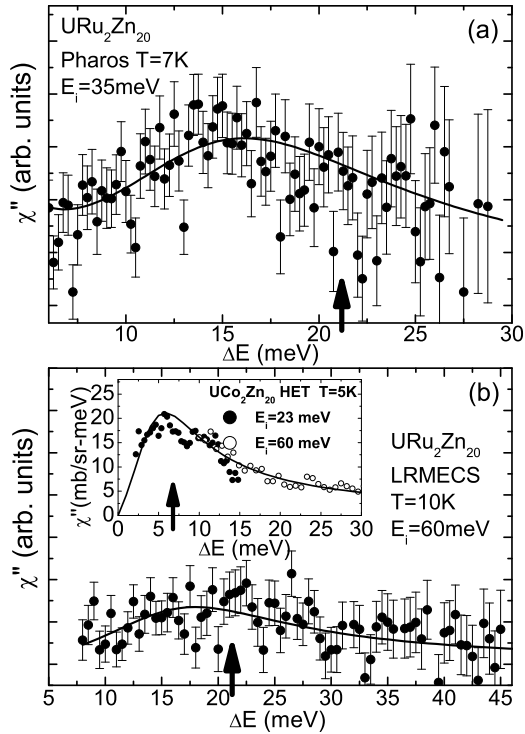}
\caption{\label{fig:2} Low temperature dynamic susceptibility
$\chi^{\prime\prime}$ vs $\Delta E$ of URu$_2$Zn$_{20}$. Error bars
in all figures represent $\pm \sigma$.(a) Pharos data at T=7 K
($E_i$ = 35 meV). (b) LRMECS data at T=10 K ($E_i$ = 60 meV). The
lines represent Lorentzian fits with  $E_0$=13.5 meV$\pm$ 1.9 meV
and $\Gamma$= 9.5 meV $\pm$ 0.6 meV. Inset: low temperature dynamic
susceptibility of UCo$_2$Zn$_{20}$; the data are from Bauer $et$
$al$\cite{BauerCo}. The line is a fit to a Lorentzian with $E_0$=3
meV $\pm$ 1.2 meV and $\Gamma$= 5 meV $\pm$ 0.4 meV. The arrows
indicate the peak positions predicted by the AIM for $N_J$ = 10 (See
Table I). } \vspace*{-3.5mm}
\end{figure}

Given that $\gamma(2K)_{Co} / \gamma(2K)_{Ru} =$ 2.6 (alternatively
$\gamma(T_{max})_{Co} / \gamma(2K)_{Ru}=$ 2.9), that $\chi(2K)_{Co}
/ \chi(2K)_{Ru}$ = 3.3, and that $E_{max}(Ru) / E_{max}(Co)$ = 2.8,
we see that at low temperature these compounds exhibit a
factor-of-three scaling of $\chi$, $\gamma$, and $E_{max}$ to an
accuracy of about 10$\%$.

\begin{figure}[t]
\centering
\includegraphics[width=0.4\textwidth]{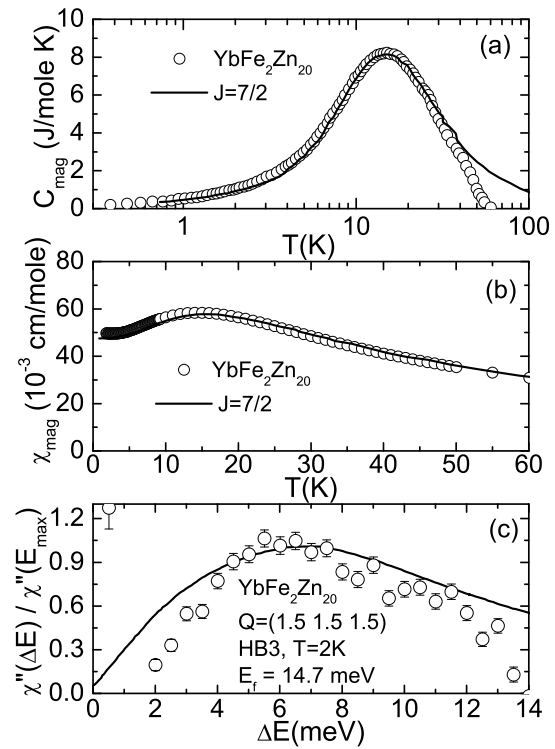}
\caption{\label{fig:3} (a) Specific heat $C_{mag}(T)$ (from Torikachvili $et$ $al$\cite{CanfieldPNAS})
and (b) magnetic susceptibility $\chi_{mag}(T)$ for YbFe$_2$Zn$_{20}$.
(c) The dynamic susceptibility $\chi^{\prime\prime} (\Delta E)/\chi^{\prime\prime} (E_{max})$ determined at
the (3/2, 3/2, 3/2) zone boundary point. The lines are fits, for the $J$ = 7/2 case,  to Rajan's predictions
for $C_{mag}$ and $\chi_{mag}$ and to Cox's predictions for $\chi^{\prime\prime} (\Delta E)/\chi^{\prime\prime}(E_{max})$.
In all three cases, there is only one common adjustable parameter $T_0$, set to the value 69 K to give the best agreement with experiment.
} \vspace*{-3.5mm}
\end{figure}

We next examine whether such scaling arises due to the applicability of the AIM to these actinide compounds. Before
doing so, we first check the validity of the AIM for the rare earth 4$f$ compound YbFe$_2$Zn$_{20}$. We apply
Rajan's Coqblin-Schrieffer model\cite{Rajan}, which is essentially the AIM in the Kondo limit ($n_{f} \simeq$ 1)
for large orbital degeneracy. In Fig. 3, we compare the data for $C_{mag}(T)$ and $\chi_{mag} (T)$ (where the
data for LuFe$_2$Zn$_{20}$ has been subtracted to determine the magnetic contribution) to Rajan's predictions
for the $J$=7/2 case\cite{Rajan}. In these fits, the only adjustable parameter is a scaling parameter $T_0$;
we find that the value 69 K gives the best agreement with experiment.

To fit to the dynamic susceptibility
$\chi^{\prime\prime} (\Delta E)$ we use the results of Cox $et$ $al$\cite{Cox}, obtained using the noncrossing
approximation (NCA) to the AIM. This calculation, which was performed for the $J$ = 5/2 case, gives the peak
position of the dynamic susceptibility at low temperature as  $E_{max}$ = 1.36 $k_B T^{Cox}_0$(see Fig. 5
in Cox $et$ $al$\cite{Cox}). The scaling temperature $T^{Cox}_0$ is related to Rajan's scaling temperature $T_0$
via $T^{Cox}_0$ = $T_0$ / 1.15 (see the caption of Fig. 2 in Cox $et$ $al$\cite{Cox}). Hence for the $J =$ 5/2
case, we have $E_{max}$ = 1.18 $k_B T_0$. In the absence of comparable theoretical results for other values of $J$,
we will assume that this relationship between $T_0$ and $E_{max}$ is approximately true for the $J =$ 7/2 and 9/2
cases; the error is probably of order 20$\%$. For YbFe$_2$Zn$_{20}$, we then expect $E_{max}$ = 7 meV.
The lineshape for $\chi^{\prime\prime}(\Delta E)/\chi^{\prime\prime}(E_{max})$
was determined from Fig. 4 in Cox $et$ $al$\cite{Cox} using this value for $E_{max}$. Albeit we have only determined
$\chi^{\prime\prime}(\Delta E)$ at one location in the zone, it is clear from these plots that the $N_J$ = 8 AIM
in the Kondo limit does an excellent job of fitting the susceptibility $\chi (T)$, magnetic specific heats
$C_{mag}$, and characteristic energy $E_{max}$ of this rare earth compound.

\begin{figure*}[t]
\centering
\includegraphics[width=0.7\textwidth]{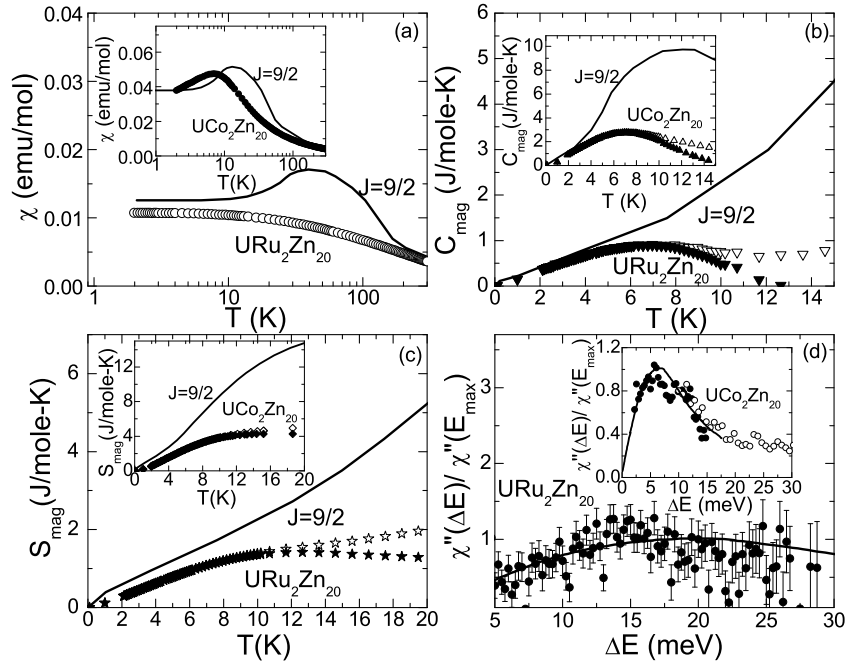}
\caption{\label{fig:4} (a) Magnetic susceptibility $\chi(T)$, (b) magnetic specific heat
$C_{mag}(T)$, and (c) entropy $S_{mag}(T)$ for URu$_2$Zn$_{20}$; the insets show the same
quantities for UCo$_2$Zn$_{20}$\cite{BauerCo}. The lines
are fits using Rajan's predictions for J=9/2. The open symbols in (b) and (c) represent data
corrected for the energy of the Einstein mode at 8 (7) meV in the U (Th) compounds\cite{endnote3}.
(d): The dynamic susceptibility $\chi^{\prime\prime}(\Delta E)/\chi^{\prime\prime}(E_{max})$ of URu$_2$Zn$_{20}$;
the inset shows the data for UCo$_2$Zn$_{20}$. The lines are obtained using Cox's results, as explained in
the text.
} \vspace*{-3.5mm}
\end{figure*}

Turning now to the actinide compounds, we note that Rajan's calculations\cite{Rajan}
for a 2J+1 Kondo impurity give the following zero-temperature limits for the
specific heat, and magnetic susceptibility:\\

$\gamma_0 = \pi J R/3T_0$\\

$\chi_0 = (2J+1)C_J/2 \pi T_0$\\

where $R$ is the gas constant and $C_J$ is the Curie constant. To
test these scaling laws, we first note that uranium has a possible
5$f^3$ state for which $J=$ 9/2 and $\mu_{eff}$=3.62$\mu_B$ ($C_J$=
1.64 emu K/mole) or a possible 5$f^2$ state for which $J=$ 4 and
$\mu_{eff}$=3.58$\mu_B$ ($C_J$= 1.60 emu K/mole). The high
temperature Curie-Weiss fit of $\chi (T)$ for URu$_2$Zn$_{20}$ gives
an experimental value for the Curie constant close to these free ion
values. In what follows, we choose $J =$ 9/2, but we note that the
analysis is not significantly different for the $J =$ 4 case.  We
estimate $T_0$ from the low temperature value for $\gamma$, and then
determine $\chi_0$. To estimate $E_{max}$ we use the above-stated
rule $E_{max}$ = 1.18 $T_0$, which as mentioned we expect to be
correct here to 20\%. The calculated results are listed in Table I,
along with the similar results for $J$ = 5/2 and $J$ = 1/2.

\begin{table*}[htp]
\caption{\label{tab:table} Experimental and theoretical values of key quantities for URu$_2$Zn$_{20}$ and UCo$_2$Zn$_{20}$.
The values for the scaling temperature $T_0$ are obtained using $\gamma_{2K}$ = 188 mJ/mol-K$^2$ for URu$_2$Zn$_{20}$ and
$\gamma_{max}$ = 558 mJ/mol-K$^2$ for UCo$_2$Zn$_{20}$. For J=9/2 and 5/2, the Curie constant
used in the calculation is the 5$f^3$ free ion value while for J=1/2, $C_J$ is obtained from
the Curie-Weiss fit to the low temperature magnetic susceptibility. }
\begin{ruledtabular}
\begin{tabular}[b]{ccccccccccc}
 &\multicolumn{2}{c}{$T_0$(K)}&\multicolumn{2}{c}{$T_{max}^C$(K)}&\multicolumn{2}{c}{$\chi_0(\frac{emu}{mole})$}&\multicolumn{2}{c}{$T_{max}^{\chi}(K)$}&\multicolumn{2}{c}{$E_{max}$(meV)}\\
  & Ru & Co & Ru & Co & Ru  & Co & Ru & Co & Ru & Co  \\
\hline
experiment &  &  & 6.8 & 7.1 & 0.0111 & 0.0372 & &7.0 &16.5 & 5.8\\
J=9/2 & 208 & 70 & 36.5 & 12.1 & 0.0125 & 0.0378 & 39.2& 13.0& 21.3&7.1\\
J=5/2 & 116 & 39 & 34.4 & 11.3 & 0.0135 & 0.0412 & 30.3 & 10.2 & 11.9& 3.9\\
J=1/2 & 23 & 7.8 & 20 & 6.8 &0.0245 & 0.0402 & & & 2.4 & 0.8 \\

\end{tabular}
\vspace*{-2mm}
\end{ruledtabular}
\end{table*}

From Table I, we can see that the expected values for $\chi_0$ and
$E_{max}$ are closer to the experimental values for the $J$ = 9/2
case than for either the $J$ = 5/2 or 1/2 cases. In Fig. 4 we
compare the experimental data to the predictions (solid lines) for
the temperature dependence of $\chi(T)$ and $C_{mag}$ (where the
data for the corresponding Th compound have been subtracted to
determine the magnetic contribution\cite{endnote3}) in the $J$ = 9/2
case. For the energy dependence of $\chi^{\prime\prime}(\Delta
E)/\chi^{\prime\prime}(E_{max})$ at low temperature, we utilize the
results of Cox $et$ $al$\cite{Cox}, as outlined above. Again, there
is only one adjustable parameter, $T_0$, which is determined from
the low temperature specific heat coefficient as equal to 208 K for
the Ru case and 70 K for the Co case. The fitting is very poor in
several respects. First, the expected values of $T_{max}$ for both
$\chi (T)$ and $C_{mag} (T)$ are much higher than observed in the
experiment, and indeed for URu$_2$Zn$_{20}$ there is even no maximum
in the experimental curve for $\chi (T)$. Even more significant is
the fact that the experimental entropy developed below 20 K is
$\emph{much}$ smaller than expected. Indeed the experimental entropy
at 20 K is less than Rln2, which would be expected for a two-fold
degeneracy (J=1/2). However, if we attempt to fit the data assuming
$J$=1/2, we find that very small values of $T_0$ are required to
reproduce the specific heat coefficients, and hence the
characteristic energy $E_{max}$ would disagree markedly with the
experimental value (see Table I). Hence there appears to be a very
serious discrepancy between the data and the Kondo model.

In our previous paper\cite{BauerCo}, we attempted to
compare the data for UCo$_2$Zn$_{20}$ to the predictions of the AIM calculated
using the NCA. The calculation assumed mixed valence between the  $J$ = 4 and
9/2 states, and assumed that a large crystal field splitting ($\sim$ 200 meV) resulted in
a six-fold degeneracy (effective $J =$ 5/2 behavior) at low temperature. To confirm whether
such a crystal field excitation is present in these compounds, we measured URu$_2$Zn$_{20}$
and ThCo$_2$Zn$_{20}$ on Pharos using large incident energies. In Fig. 5(a) we
show the  INS spectra for energy transfers up to $\Delta E$ = 550 meV. The results
exhibit no sign of crystal field excitations. We believe that a similar result will be
valid for UCo$_2$Zn$_{20}$. Furthermore, it is clear from Table I that
such an effective $J=$ 5/2 approach will overestimate $T_{max}^C$, underestimate
$E_{max}$ and badly overestimate the entropy so that the use of the AIM to describe
this compound is problematic.

Hence, while the $J$ = 7/2 AIM works extremely well
\cite{CanfieldPNAS} for the susceptibility and specific heat and
also reproduces the characteristic energy $E_{max}$ of the neutron
spectrum of YbFe$_2$Zn$_{20}$, for these actinide compounds, the J =
9/2 (or $J =$ 4) AIM works well only for the low temperature
scaling, but very poorly for the overall temperature dependence of
$\chi (T)$ and $C(T)$; in particular the theory badly overestimates
the entropy. For calculations based on smaller values of $N_J$, the
characteristic energy $E_{max}$ is badly underestimated by the
theory. These results suggest that the physics responsible for the
low temperature heavy mass behavior in these actinide compounds is
not that of local moments subject to the Kondo effect, as for the
4$f$ electron compounds, but is that of itinerant 5$f$ electrons
subject to correlation enhancement. In support of this, we note that
when uranium compounds such as UPd$_3$ exhibit local moments, then
intermultiplet excitations can be observed at energies near 400 mev;
no such excitation is seen for metallic compounds such as
UPt$_3$\cite{Osborn}. The lack of such excitations in the Pharos
data (Fig. 5(a)) for URu$_2$Zn$_{20}$ gives further evidence that
the 5$f$ electrons are itinerant, not localized, in these compounds.

\begin{figure}[t]
\centering
\includegraphics[width=0.45\textwidth]{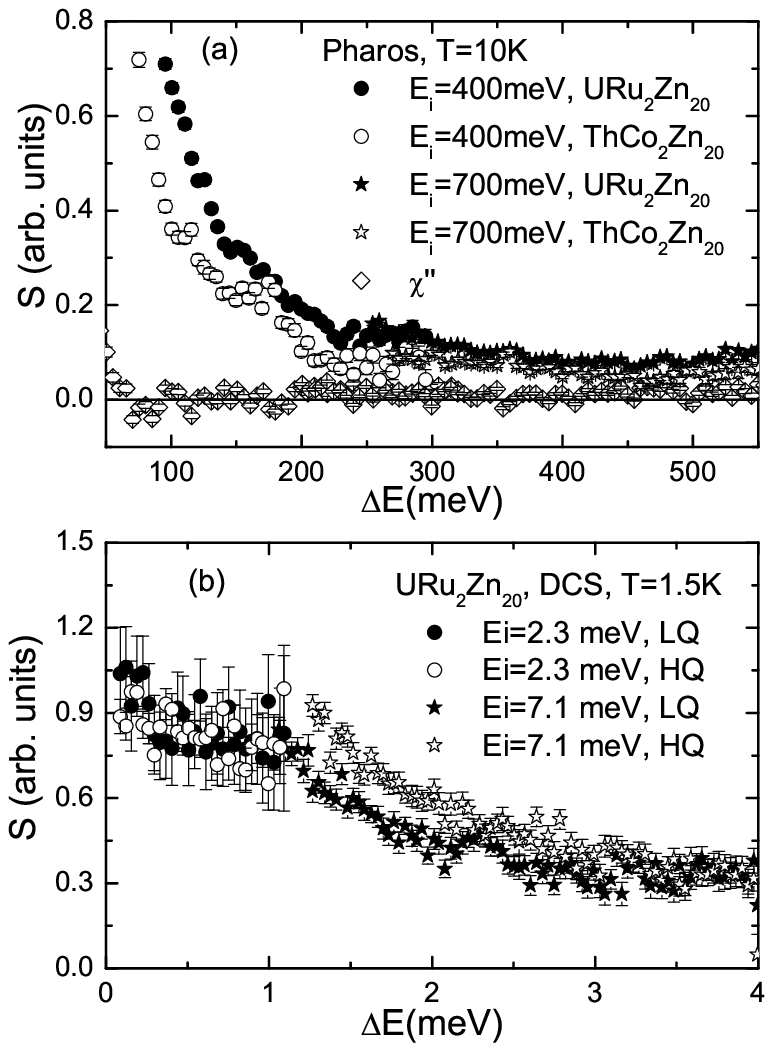}
\caption{\label{fig:5} (a) The INS spectra of URu$_2$Zn$_{20}$ and ThCo$_2$Zn$_{20}$
taken on Pharos with incident energies $E_i$ = 400 meV and 700 meV.  The diamond is the
estimated magnetic scattering $\chi^{\prime\prime}$, obtained as described in the text.
(b) The INS spectra of URu$_2$Zn$_{20}$ in the energy range 0.1-4 meV taken on DCS with
incident energies $E_i$ = 2.3 meV and 7.1 meV. The near equality of the high-$Q$ and
low-$Q$ scattering suggests that all the scattering observed in this energy range
is due to background.
} \vspace*{-3.5mm}
\end{figure}

Since the peaks observed in $C_{mag} (T)$ for both the Ru and Co
cases and in $\chi(T)$ for the Co case occur at a much lower
temperature than the characteristic temperature $E_{max}/k_B$, they
are very probably associated with low temperature magnetic
correlations, which exist only in the vicinity of some critical
wavevector $Q_N$, and which yield only a fraction of Rln2 for the
entropy. In this regard, the behavior is similar to that of
UBe$_{13}$ or UPt$_3$, where $Q$-dependent antiferromagnetic
fluctuations occur on a much smaller energy scale ($\sim$ 1 meV for
UBe$_{13}$ and 0.2 meV for UPt$_3$) than the scale of the
Kondo-like\cite{endnote2} fluctuations (13 meV for UBe$_{13}$ and 6
meV for UPt$_3$\cite{Lander,UPt3}). Such antiferromagnetic
fluctuations are large only in the vicinity of the wavevector $Q_N$
and contain only a small fraction of the spectral weight compared to
the Kondo-like fluctuations. Hence, it is not surprising that the
polycrystalline averaged INS spectra in Fig. 5(b) shows no obvious
excitation in the energy transfer range 0.1 meV to 4 meV; careful
measurements on single crystals are required to reveal such low
energy, low-spectral-weight $Q$-dependent magnetic fluctuations.

Given these considerations, we believe the characteristic INS
energies of 5.8 and 16.5 meV that we have observed in
UCo$_2$Zn$_{20}$ and URu$_2$Zn$_{20}$ represent
Kondo-like\cite{endnote2} fluctuations as observed in many uranium
compounds such as UBe$_{13}$\cite{Lander}, UPt$_3$\cite{UPt3} and
USn$_3$\cite{USn3}. The small magnetic entropy remains a difficulty,
however, even for this case. To see this, consider the scaling
product $\gamma E_{max}/k_B$, which represents how the $T$-linear
entropy is generated by the damped spin excitation centered at
$E_{max}$. For a Kondo ion, this product takes the value $\pi J
R/3$. A crude approximation would be $\gamma E_{max}/k_B$ = 2$R
ln(2J+1)$, which might be expected to be valid even for spin
fluctuations arising in an itinerant electron system; this
approximation gives a similar value ($\sim$ 39) for the $J =$ 9/2
case. The measured values for UCo$_2$Zn$_{20}$ and URu$_2$Zn$_{20}$
are in the range 33-37, very close to the expected $J =$ 9/2 value.
Fig. 4 indicates, however, that the compounds generate entropy in a
manner that satisfies this formula only at the lowest temperatures,
but then saturate above 10K. The point is that if the scaling
product has the right value, then the $R ln(2J+1)$ entropy should
continue to be generated up to temperatures of order $E_{max}/k_B$,
much larger than 10 K for these compounds. We emphasize that this
should be true even for itinerant 5$f$ electrons.

\section{conclusion}

The static and dynamic magnetic susceptibility and the specific heat
of URu$_2$Zn$_{20}$ and YbFe$_2$Zn$_{20}$ compounds have been
presented. The results show that the AIM works very well to describe
the magnetic susceptibility, specific heat and dynamic
susceptibility well of the compound YbFe$_2$Zn$_{20}$ where the 4$f$
electrons are localized. In the actinide compounds
URu$_2$Zn$_{20}$(UCo$_2$Zn$_{20}$), however, the fits to the AIM
temperature dependence are very poor even though the low temperature
scaling behavior expected for a $J =$ 9/2 Kondo impurity was
observed. An associated problem is that the magnetic entropy
generated by 20 K is too small compared to the expected value. These
results suggest that the spin fluctuations in these actinide
compounds arise from itinerant rather than localized 5$f$ electrons.
Antiferromagnetic fluctuations may affect the specific heat. While
our neutron scattering results for a polycrystalline sample saw no
signs of these fluctuations in the 0.1 to 4 meV range, they may be
observable as a small spectral weight signal in single crystal
experiments.

\section{acknowledgments}

Research at UC Irvine was supported by the U.S. Department of
Energy, Office of Basic Energy Sciences, Division of Materials
Sciences and Engineering under Award DE-FG02-03ER46036. Work at ORNL
was supported by the Scientific User Facilities Division Office of
Basic Energy Sciences (BES), DOE. Work at ANL was supported by
DOE-BES under contract DE-AC02-06CH11357. Work at the Ames
Laboratory was supported by the DOE-BES under Contract No.
DE-AC02-07CH11358. Work at Los Alamos, including work performed at
the Los Alamos Neutron Science Center, was also supported by the
DOE-BES. Work at NIST utilized facilities supported in part by the
National Science Foundation under Agreement NO. DMR-0454672.


\begin{thebibliography}{}
\bibitem{endnote}
Alhough the coherent approximation ($Q^2$ dependence) is not
expected to be valid for a single crystal, the resulting magnetic
scattering in this single crystal is identical to that obtained in
polycrystals ( A. D. Christianson et al. unpublished data) where the
approximation is valid. This may reflect the very complex unit cell
for this compound which includes 184 atoms/unit cell.
\bibitem{endnote3} In our recent paper\cite{BauerCo} we found that the optic phonon peak seen
at 8 meV in UCo$_2$Zn$_{20}$ occurs at 7 meV in ThCo$_2$Zn$_{20}$.
This shift arises from the contribution of the 5$f$ electrons to the
crystal bonding. The specific heat for this optic mode will thus be
larger for the latter compound, causing an error at low temperature
in the magnetic specific heat obatained by the U - Th subtraction.
The open symbols in Fig. 4 have corrected for this effect, assuming
Einstein modes with three degrees of freedom. The effect is small
for $T <$ 20K, hence the small magnetic entropy cannot be explained
on this basis.
\bibitem{endnote2} We call these fluctuations "Kondo-like" because they represent excitation of
the magnetic moment at temperatures $T > E_{max}/k_B$, but where the moment arises from
itinerant rather than local electrons.
\end{thebibliography}
\end{document}